\documentclass[natbib]{svjour3}
\usepackage{graphicx}
\usepackage{mathptmx} 
\begin{document}

\title{Statistical-Mechanical Measure of Stochastic Spiking Coherence in A
       Population of Inhibitory Subthreshold Neurons}

\author{Woochang Lim \and Sang-Yoon Kim}

\institute{Woochang Lim \and Sang-Yoon Kim (corresponding author) \at
           Department of Physics, Kangwon National University,
           Chunchon, Kangwon-Do 200-701, Korea \\
           \email{sykim@kangwon.ac.kr}
}

\date{Received: date / Accepted: date}

\maketitle

\begin{abstract}
By varying the noise intensity, we study stochastic spiking coherence (i.e., collective coherence between noise-induced neural spikings) in an inhibitory population of subthreshold neurons (which cannot fire spontaneously without noise). This stochastic spiking coherence may be well visualized in the raster plot of neural spikes. For a coherent case, partially-occupied ``stripes"
(composed of spikes and indicating collective coherence) are formed in the raster plot. This partial occupation occurs due to ``stochastic spike skipping'' which is well shown in the multi-peaked interspike interval histogram. The main purpose of our work is to quantitatively measure the degree of stochastic spiking coherence seen in the raster plot. We introduce a new spike-based coherence measure $M_s$ by considering the occupation pattern and the pacing pattern of spikes in the stripes. In particular, the pacing degree between spikes is determined in a statistical-mechanical way by quantifying the average contribution of (microscopic) individual spikes to the (macroscopic) ensemble-averaged global potential. This ``statistical-mechanical'' measure $M_s$ is in contrast to the conventional measures such as the ``thermodynamic'' order parameter (which concerns the time-averaged fluctuations of the macroscopic global potential), the ``microscopic'' correlation-based measure (based on the cross-correlation between the microscopic individual potentials), and the measures of precise spike timing (based on the peri-stimulus time histogram). In terms of $M_s$, we quantitatively characterize the stochastic spiking coherence, and find that $M_s$ reflects the degree of collective spiking coherence seen in the raster plot very well. Hence, the ``statistical-mechanical'' spike-based measure $M_s$ may be used usefully to
quantify the degree of stochastic spiking coherence in a statistical-mechanical way.

\end{abstract}

\section{Introduction}
\label{sec:INT}

Recently, brain rhythms have attracted  much attention \citep{Buz}. Synchronous oscillations in neural systems may be used for efficient sensory processing (e.g., binding of the integrated whole image in the visual cortex is accomplished via synchronization of neural firings) \citep{Gray}. In addition to such neural encoding of sensory stimuli, neural synchronization is also correlated with pathological rhythms associated with neural diseases (e.g., epileptic seizures and tremors in the Parkinson's disease) \citep{ND1,ND2,ND3}. Here, we are interested in these synchronized neural oscillations.

A neural circuit in the major parts of the brain such as thalamus, hippocampus, and cortex consists of a few types of excitatory principal cells and diverse types of inhibitory interneurons. Functional diversity of interneurons increases the computational power of principal cells \citep{Buz,Inhibitory}. To understand the mechanisms of synchronous brain rhythms, neural systems composed of excitatory neurons and/or inhibitory neurons have been investigated, and thus three types of synchronization mechanisms have been found \citep{Wang}:
recurrent excitation between principal cells, mutual inhibition between interneurons, and feedback between excitatory and inhibitory neurons. Perfect synchronization occurs in the network of pulse-coupled excitatory neurons \citep{ME1,ME2}. However, this synchronization via mutual excitation cannot be always stable in networks with slowly decaying synaptic couplings. Stability of synchronization is shown to depend on both the time course of synaptic interaction and the response of neurons to small depolarization \citep{Stab1,Stab2}. When the decay time of the synaptic interaction is enough long, mutual inhibition (rather than excitation) may synchronize neural firing activities. By providing a coherent oscillatory output to the principal cells, interneuronal networks play the role of the backbones (i.e., synchronizers or pacemakers) of many brain oscillations such as the 10-Hz thalamocortical spindle rhythms \citep{WR,GR} and the 40-Hz fast gamma rhythms in the hippocampus and the neocortex \citep{WB,White,ING,Gamma1}. When the feedback
between the excitatory and inhibitory populations is strong, neural synchrony appears via the ``cross-talk'' between the two populations \citep{ING,Gamma1,HM,BK1,BK2}.

Most past studies exploring mechanisms of neural synchronization were done in neural systems composed of spontaneously firing (i.e., self-oscillating) suprathreshold neurons. For this case, neural coherence occurs via cooperation of regular firings of suprathreshold self-firing neurons. In contrast, neural systems composed of subthreshold neurons have received little attention. Unlike the suprathreshold case, each subthreshold neuron in the absence of coupling cannot fire spontaneously without noise; it can fire only with the help of noise. Recently, stochastic spiking coherence (i.e., collective coherence emerging via cooperation of noise-induced spikings) was observed in an excitatory population of pulse-coupled subthreshold neurons \citep{CR,Kim1,Kim2}. This kind of works may be thought to correspond to
a ``subthreshold version'' of neural synchronization through mutual excitation. Due to the stochastic spiking coherence, synaptic current, injected into each
individual neuron, becomes temporally coherent. Hence, temporal coherence resonance of an individual subthreshold neuron in the network may be enhanced. This enhancement of coherence resonance in an excitatory network of subthreshold Hodgkin-Huxley neurons was characterized in terms of the coherence factor $\beta$, representing the degree of sharpness of the peak in the power spectrum of an individual neuron \citep{CR}. In this way, the measure $\beta$ is used for characterization of the temporal coherence of an individual neuron. However, we note that $\beta$ is not a measure for directly measuring the degree of collective coherence in the whole population.

In this paper, we are concerned about the ``subthreshold version'' of neural
synchronization via mutual inhibition. In Section \ref{sec:ML}, we describe the biological conductance-based Morris-Lecar (ML) neuron model with voltage-gated
ion channels \citep{ML1,ML2,ML3}. The ML neurons (used in our study) exhibit the type-II excitability (i.e., the firing frequency begins to increase from a
non-zero value when the stimulus exceeds a threshold value), and they interact
via inhibitory GABAergic synapses whose activity increases fast and decays slowly. In Section \ref{sec:SSC}, we characterize stochastic spiking coherence in a large population of inhibitory subthreshold ML neurons by varying the noise intensity for a fixed coupling strength. Weakly coherent states with
oscillating ensemble-averaged global membrane potential $V_G$ are thus found
to appear in a range of intermediate noise intensity [i.e., regular global
oscillation (with reduced amplitude and increased frequency) emerges via
cooperation of irregular individual oscillations]. Emergence of collective
coherence may be well described in terms of the conventional ``thermodynamic''
order parameter which concerns the time-averaged fluctuations of the macroscopic global potential $V_G$. We note that this stochastic spiking coherence may be well visualized in the raster plot of neural spikes (i.e., a spatiotemporal plot of neural spikes) which is directly obtained in experiments. For the coherent case, ``stripes" (composed of spikes and indicating collective coherence) are found to be formed in the raster plot. Due to coherent contribution of spikes, local maxima of the global potential $V_G$ appear at the centers of stripes. However, these stripes are partially occupied. Individual inhibitory neurons exhibit intermittent spikings phase-locked to $V_G$ at random multiples of the period of $V_G$. This ``stochastic phase locking'' leading to ``stochastic spike skipping''
is well shown in the interspike interval (ISI) histogram with multiple peaks.
The multi-peaked ISI histogram shows some indication of weak collective spiking coherence.

The main purpose of our work is to quantitatively measure the degree of stochastic spiking coherence seen in the raster plot. In Section \ref{sec:SSC}, we introduce a new type of spike-based coherence measure $M_s$ by taking into consideration the occupation pattern and the pacing pattern of spikes in the stripes of the raster plot. In particular, the pacing degree between spikes is determined in a statistical-mechanical way by quantifying the average contribution of (microscopic) individual spikes to the (macroscopic) global potential $V_G$. This ``statistical-mechanical'' measure $M_s$ is in contrast to the conventional measures such as the ``thermodynamic'' order parameter \citep{GR,HM}, the ``microscopic'' correlation-based measure (based on the cross-correlations between the microscopic individual potentials) \citep{WB,White}, and the measures of precise spike timing based on the peri-stimulus time histogram (PSTH) \citep{PSTH1,PSTH2}. The ``thermodynamic'' order parameter and the ``microscopic'' measure concern just the the macroscopic global potential $V_G$ and the microscopic individual potentials,
respectively without considering any quantitative relation between $V_G$ and the microscopic individual potentials. (The auto-correlation of the global activity used in the work of Brunel and Hakim (1999) may also be regarded as a kind of ``thermodynamic'' measure.) For the PSTH-based measure ``events,'' corresponding to peaks of the instantaneous population firing rate, are selected through setting a threshold. Then, the measures for the reliability and the precision of spike timing concern only the spikes within the events, in contrast to the case of the ``statistical-mechanical'' measure where all spikes are considered (without selecting events). A main difference between the
conventional and the new spike-based measures lies in determining the pacing degree of spikes. The precision of spike timing for the conventional case is given by just the standard deviation of (microscopic) individual spike times within an event without considering the quantitative contribution of (microscopic) individual spikes to the (macroscopic) global activity. Hence, the PSTH-based measure is not a statistical-mechanical measure. However, if we take the instantaneous population firing rate as a global activity and exactly define the global cycles [see Fig.~\ref{fig:SM1}(a)] and the global phases  [see Eqs. (\ref{eq:GP1}) and (\ref{eq:GP2})] like our case, then the conventional PSTH-based measure may also develop into a similar statistical-mechanical measure. By varying the noise intensity, we quantitatively characterize the stochastic spiking coherence in terms of the ``statistical-mechanical'' measure $M_s$, and find that $M_s$ reflects the degree of collective spiking coherence seen in the raster plot very well. We also expect that $M_s$ may be implemented for characterizing the degree of collective coherence in the experimentally-obtained raster plot of neural spikes. Finally, a summary is given in Section \ref{sec:SUM}.

\section{Morris-Lecar Neuron Model}
\label{sec:ML}

In this section, we describe the biological neuron model used in our computational study. We consider an ensemble of $N$ globally coupled neurons. As an element in our neural system, we choose the conductance-based ML neuron model with voltage-gated ion channels, originally proposed to describe the time-evolution pattern of the membrane potential for the giant muscle fibers of barnacles \citep{ML1,ML2,ML3}. The population dynamics in this neural network is governed by a set of the following differential equations:
\begin{eqnarray}
C \frac{dv_i}{dt} &=& -I_{ion,i}+I_{DC} +D \xi_{i} -I_{syn,i}, \label{eq:CML1A} \\
\frac{dw_i}{dt} &=& \phi \frac{(w_{\infty}(v_i) -
w_i)}{\tau_R(v_i)}, \label{eq:CML1B} \\
\frac{ds_i}{dt}&=& \alpha s_{\infty}(v_i) (1-s_i) -
\beta s_i, \;\;\; i=1, \cdots, N, \label{eq:CML1C}
\end{eqnarray}
where
\begin{eqnarray}
I_{ion,i} &=& I_{Ca,i} + I_{K,i} + I_{L,i} \label{eq:CML2A} \\
&=& g_{Ca} m_{\infty} (v_i) (v_i - V_{Ca}) + g_K w_i (v_i - V_K) +
g_L (v_i - V_L), \label{eq:CML2B} \\
I_{syn,i} &=& \frac{J}{N-1} \sum_{j(\ne i)}^N s_j(t) (v_i -
V_{syn}), \label{eq:CML2C} \\
m_{\infty}(v) &=& 0.5 \left[ 1+\tanh \left\{ (v-V_1)/{V_2}
\right\} \right], \label{eq:CML2D} \\
w_{\infty}(v)&=& 0.5 \left[ 1+\tanh \left\{
(v-V_3)/{V_4} \right\} \right], \label{eq:CML2E} \\
\tau_{R}(v) &=& 1/ \cosh \left\{ (v-V_3)/(2V_4) \right\}, \label{eq:CML2F} \\
s_{\infty} (v_i) &=& 1/[1+e^{-(v_i-v^*)/\delta}]. \label{eq:CML2G}
\end{eqnarray}
Here, the state of the $i$th neuron at a time $t$ (measured in units of ms) is characterized by three state variables: the membrane potential $v_i$ (measured in units of mV), the slow recovery variable $w_i$ representing the activation of the $K^+$ current (i.e., the fraction of open $K^+$ channels), and the synaptic gate variable $s_i$ denoting the fraction of open synaptic ion channels. In Eq.~(\ref{eq:CML1A}), $C$ represents the capacitance of the
membrane of each neuron, and the time evolution of $v_i$ is governed by four kinds of source currents.

The total ionic current $I_{ion,i}$ of the $i$th neuron consists of the calcium current $I_{Ca,i}$, the potassium current $I_{K,i}$, and the leakage current $I_{L,i}$. Each ionic current obeys Ohm's law. The constants $g_{Ca}$, $g_{K}$, and $g_{L}$ are the maximum conductances for the ion and leakage channels, and the constants $V_{Ca}$, $V_K$, and $V_L$ are the reversal potentials at which each current is balanced by the ionic concentration difference across the membrane. Since the calcium current $I_{Ca,i}$ changes much faster than the potassium current $I_{K,i}$, the gate variable $m_i$ for the $Ca^{2+}$ channel is assumed to always take its saturation value $m_\infty(v_i)$. On the other hand, the activation variable $w_i$ for the $K^{+}$ channel approaches its
saturation value $w_{\infty}(v_i)$ with a relaxation time $\tau_R(v_i) / \phi$, where $\tau_R$ has a dimension of ms and $\phi$ is a (dimensionless) temperature-like time scale factor.

Each ML neuron is also stimulated by the common DC current $I_{DC}$ and an independent Gaussian white noise $\xi$ [see the 2nd and 3rd terms in Eq.~(\ref{eq:CML1A})] satisfying $\langle \xi_i(t) \rangle =0$ and $\langle \xi_i(t)~\xi_j(t') \rangle = \delta_{ij}~\delta(t-t')$, where $\langle\cdots\rangle$ denotes the ensemble average. The noise $\xi$ is a parametric one which randomly perturbs the strength of the applied
current $I_{DC}$, and its intensity is controlled by the parameter $D$. The last term in Eq.~(\ref{eq:CML1A}) represents the coupling of the network. Each neuron is connected to all the other ones through global synaptic couplings. $I_{syn,i}$ of Eq.~(\ref{eq:CML2C}) represents such synaptic current injected into the $i$th neuron. Here the coupling strength is controlled by the parameter $J$ and $V_{syn}$ is the synaptic reversal potential. We use $V_{syn}=-80$ mV for the inhibitory synapse and $V_{syn}=0$ mV for the excitatory synapse. The synaptic gate variable $s$ obeys the 1st order kinetics of Eq.~(\ref{eq:CML1C}) \citep{GR,WB}. Here, the normalized concentration of synaptic transmitters, activating the synapse, is assumed to be an instantaneous sigmoidal function of the membrane potential with a threshold $v^*$ in Eq.~(\ref{eq:CML2G}), where we set $v^*=0$ mV and $\delta=2$ mV. The transmitter release occurs only when the neuron emits a spike (i.e., its potential $v$ is larger than $v^*$). The synaptic channel opening rate, corresponding to the inverse of the synaptic rise time $\tau_r$, is $\alpha=10$ ${\rm ms}^{-1}$, which assures a fast rise of $I_{syn}$ \citep{BK1,BK2}. On the other hand, the synaptic closing rate $\beta$, which is the inverse of the synaptic decay time $\tau_d$, depends on the type of the synaptic receptors. For the inhibitory GABAergic synapse (involving the $\rm {GABA_A}$ receptors) we set $\beta=0.1$ ${\rm ms}^{-1}$, and for the excitatory glutamate synapse (involving the AMPA receptors), we use $\beta=0.5$ ${\rm ms}^{-1}$
\citep{BK1,BK2}. Thus, $I_{syn}$ decays slowly.

The ML neuron may exhibit either type-I or type-II excitability, depending on the system parameters \citep{ML2}. For the case of type-I (type-II), the firing frequency begins to increase from zero (a non-zero finite value) when $I_{DC}$ passes a threshold. Throughout this paper, we consider the case of type-II excitability where $g_{Ca} = 4.4~ {\rm mS/cm^2},\, g_{K} = 8~ {\rm mS/cm^2},\, g_{L} = 2~{\rm mS/cm^2},$ $V_{Ca} = 120~ {\rm mV},\, V_{K}=-84~ {\rm mV},\,V_{L} = -60~ {\rm mV},$ $C = 20~ \mu {\rm F/cm^2},$ $\phi = 0.04,$ $V_1 = -1.2~ {\rm mV},\, V_2 = 18~ {\rm mV},\, V_3 = 2~{\rm mV},$ and $V_4 = 30~ {\rm mV}$. As $I_{DC}$ passes a threshold in the absence of noise, each single type-II ML neuron begins to fire with a nonzero frequency that is relatively insensitive to the change in $I_{DC}$ \citep{Excitability1,Excitability2}.
Numerical integration of Eqs.~(\ref{eq:CML1A}-\ref{eq:CML1C}) is done using the Heun method \citep{SDE} (with the time step $\Delta t=0.01$ ms) similar to the second-order Runge-Kutta method, and data for $(v_i,w_i,s_i)$ $(i=1,\dots,N)$ are obtained with the sampling time interval $\Delta t=1$ ms. For each realization of the stochastic process in Eqs.~(\ref{eq:CML1A}-\ref{eq:CML1C}), we choose a random initial point $[v_i(0),w_i(0),s_i(0)]$ for the $i$th $(i=1,\dots, N)$ neuron with uniform probability in the range of $v_i(0) \in (-70,50)$, $w_i(0) \in (0.0,0.6)$, and $s_i(0) \in (0.0,1.0)$.

\section{Characterization of Stochastic Spiking Coherence in An Inhibitory
Population of Subthreshold ML Neurons}
\label{sec:SSC}

In this section, we study stochastic spiking coherence in a population of inhibitory subthreshold ML neurons. The stochastic spiking coherence is quantitatively characterized in terms of a new ``statistical-mechanical'' spike-based measure.

\begin{figure}[t]
\centerline{\includegraphics[width=0.8\columnwidth]{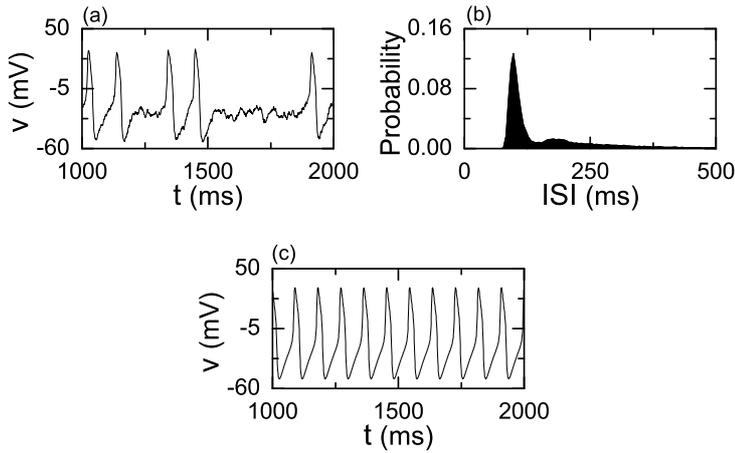}}
\caption{Intermittent noise-induced firings of a single subthreshold  ML neuron: (a) time series of the membrane potential v and (b) interspike interval (ISI) histogram for $I_{DC}=87$ $\mu {\rm A /cm^2}$ and $D=20$ ${\rm \mu A \cdot {ms}^{1/2}/cm^2}$; the ISI histogram is made of $5 \times 10^4$ ISIs and the bin size for the histogram is 5 ms. Regular firings of a single suprathreshold ML neuron: (c) time series of v for $I_{DC}=95$ $\mu {\rm A /cm^2}$ and $D=0$.}
\label{fig:SN}
\end{figure}

We first consider the case of a single ML neuron. An isolated subthreshold neuron cannot fire spontaneously without noise; it may generate firings only with the aid of noise. Figure \ref{fig:SN}(a) shows a time series of the membrane potential $v$ of a subthreshold neuron for $I_{DC}=87$ $\mu {\rm A /cm^2}$ and $D=20$ ${\rm \mu A \cdot {ms}^{1/2}/cm^2}$. Noise-induced spikings appear intermittently. For this subthreshold case, the ISI histogram is
shown in Fig.~\ref{fig:SN}(b). The most probable value of the ISIs (corresponding to the main highest peak) is $97.5$ ms. But, due to a long tail in the ISI histogram the average value of the ISIs becomes 161.6 ms. This noise-induced irregular oscillation of $v$ is in contrast to the regular oscillation of $v$ for a suprathreshold neuron (which fires spontaneously without noise) when $I_{DC}=95$ $\mu {\rm A /cm^2}$ and $D=0$ [see Fig.~\ref{fig:SN}(c)].

\begin{figure}[t]
\centerline{\includegraphics[width=0.8\columnwidth]{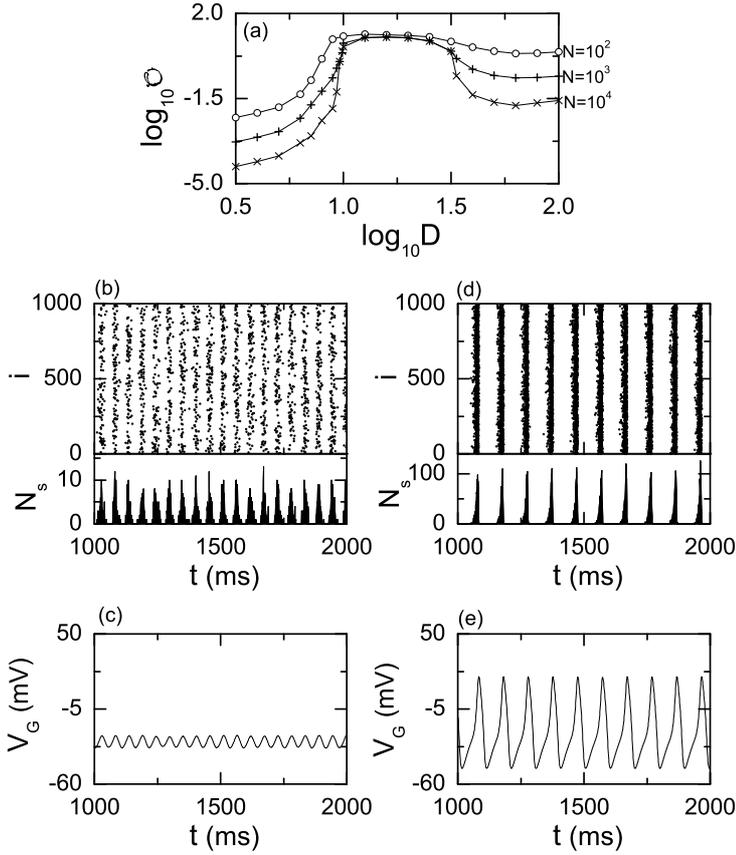}}
\caption{Order parameter and partial occupation in $N (=10^3)$ globally coupled inhibitory subthreshold ML neurons: (a) plots of the order parameter $\cal{O}$ versus the noise intensity $D$. (b) partially-occupied raster plot of spikings a ($i$: neuron index) and plot of number of spikes ($N_s$) versus time ($t$) and (c) time series of the ensemble-averaged global potential $V_G$. Full occupation in $N(=10^3)$ globally coupled excitatory ML neurons: (d) fully-occupied raster plot of spikings and plot of $N_s$ versus $t$ and (e) time series of $V_G$. For both inhibitory and excitatory cases, $I_{DC}=87$ $\mu {\rm A /cm^2}$, $D=20$ ${\rm \mu A \cdot {ms}^{1/2}/cm^2}$, $J=3$ ${\rm mS /cm^2}$, and the bin size for $N_s$ in (b) and (d) is 1 ms}
\label{fig:PO}
\end{figure}

We consider an inhibitory population of $N$ globally coupled subthreshold ML neurons for $I_{DC}=87$ $\mu {\rm A /cm^2}$ and set the coupling strength as $J=3$ ${\rm mS /cm^2}$. (Hereafter, for convenience we omit the dimensions of $I_{DC}$, $D$, and $J$.) By varying the noise intensity $D$, we investigate the stochastic spiking coherence. Emergence of collective spiking coherence may be well described by the (population-averaged) global potential,
\begin{equation}
 V_G (t) = \frac {1} {N} \sum_{i=1}^{N} v_i(t).
\label{eq:GPOT}
\end{equation}
In the thermodynamic limit $(N \rightarrow \infty)$, a collective state becomes coherent if $\Delta V_G(t)$ $(= V_G(t) - \overline{V_G(t)})$ is non-stationary (i.e., an oscillating global potential $V_G$ appears for a coherent case), where the overbar represents the time average. Otherwise (i.e., when $\Delta V_G$ is stationary), it becomes incoherent. Thus, the mean square deviation of the global potential $V_G$ (i.e., time-averaged fluctuations of $V_G$),
\begin{equation}
{\cal{O}} \equiv \overline{(V_G(t) - \overline{V_G(t)})^2},
 \label{eq:Order}
\end{equation}
plays the role of an order parameter used for describing the coherence-incoherence transition \citep{Order}. For the coherent (incoherent) state, the order parameter $\cal{O}$ approaches a non-zero (zero) limit value as $N$ goes to the infinity. Figure \ref{fig:PO}(a) shows a plot of the order parameter versus the noise intensity. For $D < D^*_l$ $(\simeq 9.4$), incoherent states exist because the order parameter $\cal{O}$ tends to zero
as $N \rightarrow \infty$. As $D$ passes the lower threshold $D^*_l$, a coherent transition occurs because of a constructive role of noise to stimulate coherence between noise-induced spikings. However, for large $D > D^*_h$ $(\simeq 33.4$) such coherent states disappear (i.e., a transition to an incoherent state occurs when $D$ passes the higher threshold $D^*_h$)
due to a destructive role of noise to spoil the collective spiking coherence.

As an example, we consider a coherent state for $D=20$. For this case stochastic spiking coherence may be well visualized in terms of the raster plot of neural spikes. The upper panel in Fig.~\ref{fig:PO}(b) shows the raster plot for $N=10^3$. We note that stripes (composed of spikes and indicating collective coherence) appear successively in the raster plot. The number of spikes $(N_s)$ in each stripe is given in the lower panel. About 100 spikes constitute a stripe [i.e., only a fraction (about 1/10) of the total
neurons fire in each stripe]. In this way partial occupation occurs in the stripes. A regularly oscillating global potential $V_G$ emerges through cooperation of spikes in partially-occupied stripes. A time series of $V_G$ is shown in Fig.~\ref{fig:PO}(c). The amplitude of $V_G$ is much smaller than that in Fig.~\ref{fig:SN}(a) for the isolated single case. This reduction of amplitude occurs mainly because of partial occupation. Local maxima of $V_G$ appear at the centers of stripes where the spiking densities are locally highest. The global period $T_G$ of $V_G$ (corresponding to the average intermax interval of $V_G$ or equivalently corresponding to the average interstripe interval in the raster plot) is 54.2 ms. Hence the global frequency $f_G$ $(=18.5$ Hz) of $V_G$ is roughly as twice as the most probable frequency $(= 10.3$ Hz) of $v$ for the isolated single case. Thus, a regular global oscillation with reduced amplitude and increased frequency occurs for the partially-occupied case. For comparison, we also consider the case of full occupation which occurs in a population of excitatory neurons
for the same set of parameters $I_{DC}$, $D$, and $J$. [We emphasize that partial occupation may also occur even for the excitatory case of smaller $J$ (e.g., $J=1$).] Figures \ref{fig:PO}(d) and \ref{fig:PO}(e) show the raster plot and a time series of $V_G$ for the fully-occupied case, respectively. Fully-occupied stripes appear successively at nearly regular time intervals $\Delta t$ ($= 97.9$ ms), in contrast to the partially-occupied case. A regularly oscillating $V_G$ with large amplitude appears via cooperation of spikes in the fully-occupied stripes, and its global frequency $f_G$ $(=10.2$ Hz) is smaller than that for the partially-occupied case (in fact, $f_G$ for the fully-occupied case is nearly equal to the most probable frequency of
$v$ for the isolated single case).

\begin{figure}[t]
\centerline{\includegraphics[width=0.8\columnwidth]{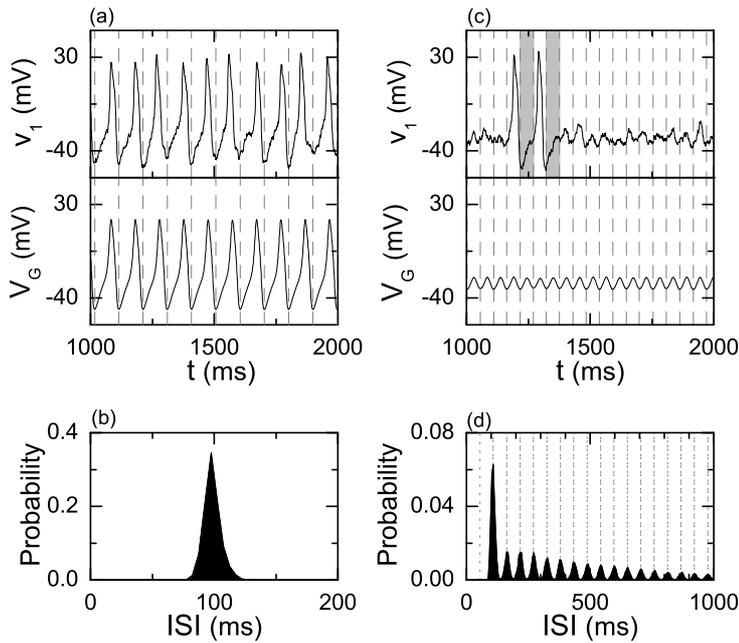}}
\caption{1:1 phase locking (spikings phase-locked to the global potential $V_G$ at every cycle of $V_G$) in $N (=10^3)$ globally coupled excitatory ML neurons: (a) time series of the local potential $v_1$ of the 1st neuron and the ensemble-averaged global potential $V_G$ and (b) interspike interval (ISI) histogram with a single peak. Stochastic phase locking leading to stochastic spike skipping (intermittent spikings phase-locked to $V_G$ at random multiples of the period of $V_G$) in $N (=10^3)$ globally coupled inhibitory subthreshold ML neurons: (c) time series of $v_1$ and $V_G$, and (d) ISI histogram with multiple peaks. For both inhibitory and excitatory cases, $I_{DC}=87$ $\mu {\rm A /cm^2}$, $D=20$ ${\rm \mu A \cdot {ms}^{1/2}/cm^2}$, and $J=3$ ${\rm mS /cm^2}$. Vertical dashed lines in (a) and (c) represent the times at which local minima of $V_G$ occur. Each ISI histogram in (b) and (d) is made of $5 \times 10^4$ ISIs and the bin size for the histogram is 5 ms. Vertical dotted lines in (d) denote integer multiples of the global period $T_G$ (=54.2 ms) of $V_G$.}
\label{fig:RSS}
\end{figure}

To further understand the full and the partial occupations in the raster plots, we examine the local and the global output signals for both the fully-occupied and the partially-occupied cases of $D=20$. Since our neural network is globally coupled, any local neuron may be a representative one. Unlike the isolated single case (shown in Fig.~\ref{fig:SN}), each local neuron within the network is stimulated by a synaptic current which is directly related to the average of firing events of all the other neurons. First, we consider the fully-occupied excitatory case, and investigate the time series of the local potential $v_1$ of the 1st neuron and the global potential $V_G$ which are shown in Fig.~\ref{fig:RSS}(a). The 1st neuron fires spikes phase-locked to $V_G$ at every cycle of $V_G$ (i.e., the local potential $v_1$ exhibits
a nearly regular phase-locked oscillation). To confirm this 1:1 phase locking, we collect $5 \times 10^4$ ISIs from all local neurons, and obtain the ISI histogram [see Fig.~\ref{fig:RSS}(b)]. A single peak with a small width is located around the average value ($=97.9$ ms) which agrees well with the global period $T_G$ of $V_G$. This 1:1 phase locking is expected to occur for the fully-occupied case because the global frequency $f_G$ $(= 10.2$ Hz) of $V_G$  is nearly equal to the most probable frequency $(= 10.3$ Hz) of noise-induced oscillation for the uncoupled single neuron. As a result of such 1:1 phase locking, full occupation occurs in the raster plot. Second, we study the partially-occupied inhibitory case. Figure \ref{fig:RSS}(c) shows the time series of the local potential $v_1$ of the 1st neuron and the global potential $V_G$. The 1st neuron exhibits intermittent spikings phase-locked to $V_G$ at random multiples of the period $T_G$ (=54.2 ms) of $V_G$. In addition to these intermittent spiking phases, hopping phases (exhibiting small subthreshold oscillations) appear in most of global cycles. We also note that after occurrence of a spiking, recovery from a hyperpolarized
to a resting state is made during the next global cycle. Hence, a ``preparatory'' phase without spiking and hopping (for preparing for generation of the next spike or hopping) follows each spiking phase (see the gray parts). In this way, the local potential exhibits an irregular firing pattern consisting of randomly phase-locked spiking and hopping phases and preparatory phases. We note that a regular fast global oscillation with a small amplitude emerges from these irregular individual oscillations. To confirm the stochastic spike skipping (arising fromn stochastic phase locking) in the local potential, we collect $5 \times 10^4$ ISIs from all local neurons and get the ISI
histogram. Multiple peaks appear at multiples of the period $T_G$ of the global potential $V_G$. However, due to appearance of preparatory cycles, the 1st peak of the histogram (which is the highest one) appears at $2\,T_G$ (not $T_G$). Hence, local neurons fire mostly in alternate global cycles. Due to this stochastic spike skipping partial occupation occurs in the raster plot. Similar skipping phenomena of spikings (characterized with multi-peaked ISI histograms) were found in single noisy neuron models driven by a weak periodic external force for the stochastic resonance \citep{SR1,Longtin,SR2}. Unlike this single case, stochastic spike skipping in networks of inhibitory subthreshold neurons is a collective effect because it occurs due to a driving by a coherent ensemble-averaged synaptic current. As discussed in details in Section \ref{sec:SUM}, similar results on skipping were also obtained in simplified
networks of IF neurons \citep{Brunel1,Brunel2,Brunel3}. However, there are some differences in the spiking pattern of individual neurons and the effect of noise on the collective coherence.

\begin{figure}[t]
\centerline{\includegraphics[width=0.8\columnwidth]{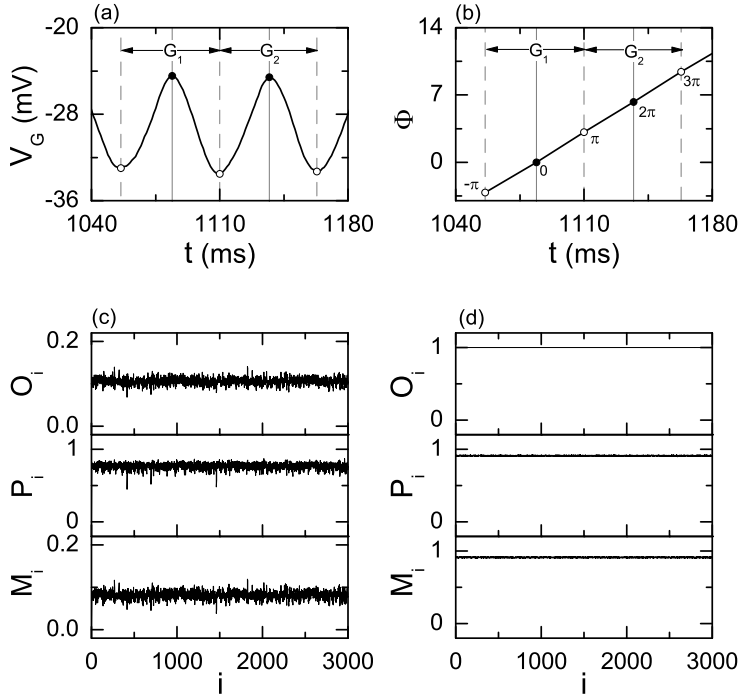}}
\caption{``Statistical-mechanical'' spike-based coherence measure in $N (=10^3)$ globally coupled ML neurons for $I_{DC}=87$ $\mu {\rm A /cm^2}$, $D=20$ ${\rm \mu A \cdot {ms}^{1/2}/cm^2}$, and $J=3$ ${\rm mS/cm^2}$. Partially-occupied inhibitory case: (a) time series of the global potential $V_G$, (b) plot of the global phase ($\Phi$) versus time ($t$), and (c) plots of $O_i$ (occupation degree of spikes in the $i$th stripe), $P_i$ (pacing degree of spikes in the ith stripe), and $M_i$ (spiking measure in the ith stripe) versus $i$ (stripe). In (a) and (b), vertical dashed and solid lines represent the times at which local minima and maxima (denoted by open and solid circles) of $V_G$ occur, respectively and $G_i$ ($i=1,2$) denotes the ith
global cycle. Fully-occupied excitatory case: (d) plots of $O_i$, $P_i$, and $M_i$ versus $i$.}
\label{fig:SM1}
\end{figure}

Our main purpose is to quantitatively characterize the stochastic spiking coherence which is well visualized in the raster plot of spikes. To measure the degree of stochastic spiking coherence seen in the raster plots in Fig.~\ref{fig:PO} for $D=20$, we introduce a new ``statistical-mechanical'' spike-based coherence measure $M_s$ by considering the occupation pattern and the pacing pattern of spikes in the stripes of the raster plot. Particularly, the pacing degree between spikes is determined in a statistical-mechanical way by quantifying the average contribution of microscopic individual spikes to the macroscopic global potential $V_G$. The spiking coherence measure $M_i$ of the $i$th stripe is defined by the product of the occupation degree $O_i$ of spikes (representing the density of the $i$th stripe) and the pacing degree
$P_i$ of spikes (denoting the smearing of the $i$th stripe):
\begin{equation}
M_i = O_i \cdot P_i.
\label{eq:SM}
\end{equation}
The occupation degree $O_i$ in the $i$th stripe is given by the fraction of spiking neurons:
\begin{equation}
   O_i = \frac {N_i^{(s)}} {N},
\end{equation}
where $N_i^{(s)}$ is the number of spiking neurons in the $i$th stripe.
For the full occupation, $O_i=1$, while for the partial occupation $O_i<1$.
The pacing degree $P_i$ of each microscopic spike in the $i$th stripe can be
determined in a statistical-mechanical way by taking into consideration its
contribution to the macroscopic global potential $V_G$. Figure \ref{fig:SM1}(a) shows a time series of the global potential $V_G$; local maxima and minima are
denoted by solid and open circles, respectively. Central maxima of $V_G$ between neighboring left and right minima of $V_G$ coincide with centers of stripes in the raster plot. The global cycle starting from the left minimum of $V_G$ which appears first after the transient time $(=10^3$ ms) is regarded as the 1st one, which is denoted by $G_1$. The 2nd global cycle $G_2$ begins from the next following right minimum of $G_1$, and so on. Then, we introduce an instantaneous global phase $\Phi(t)$ of $V_G$ via linear interpolation in the two successive subregions forming a global cycle \citep{Phase}. The global phase $\Phi(t)$ between the left minimum (corresponding to the beginning point of the $i$th global cycle) and the central maximum is given by:
\begin{equation}
\Phi(t) = 2\pi(i-3/2) + \pi \left(
\frac{t-t_i^{(min)}}{t_i^{(max)}-t_i^{(min)}} \right)
 {\rm~~ for~} ~t_i^{(min)} \leq  t < t_i^{(max)}
~~(i=1,2,3,\dots),
\label{eq:GP1}
\end{equation}
and $\Phi(t)$ between the central maximum and the right minimum (corresponding to the beginning point of the $(i+1)$th cycle) is given by:
\begin{equation}
\Phi(t) = 2\pi(i-1) + \pi \left(
\frac{t-t_i^{(max)}}{t_{i+1}^{(min)}-t_i^{(max)}} \right)
 {\rm~~ for~} ~t_i^{(max)} \leq  t < t_{i+1}^{(min)}
~~(i=1,2,3,\dots),
\label{eq:GP2}
\end{equation}
where $t_i^{(min)}$ is the beginning time of the $i$th global cycle (i.e., the time at which the left minimum of $V_G$ appears in the $i$th global cycle) and $t_i^{(max)}$ is the time at which the maximum of $V_G$ appears in the $i$th global cycle. The global phase $\Phi$ in the first two global cycles is shown in Fig.~\ref{fig:SM1}(b). Then, the contribution of the $k$th microscopic spike in the $i$th stripe occurring at the time $t_k^{(s)}$ to $V_G$ is given by $\cos \Phi_k$, where $\Phi_k$ is the global phase at the $k$th spiking time [i.e., $\Phi_k \equiv \Phi(t_k^{(s)})$]. A microscopic spike makes the most constructive (in-phase) contribution to $V_G$ when the corresponding
global phase $\Phi_k$ is $2 \pi n$ ($n=0,1,2, \dots$), while it makes the most
destructive (anti-phase) contribution to $V_G$ when $\Phi_i$ is $2 \pi (n-1/2)$. By averaging the contributions of all microscopic spikes in the $i$th stripe to $V_G$, we obtain the pacing degree of spikes in the $i$th stripe,
\begin{equation}
 P_i ={ \frac {1} {S_i}} \sum_{k=1}^{S_i} \cos \Phi_k,
\end{equation}
where $S_i$ is the total number of microscopic spikes in the $i$th stripe.
By averaging $M_i$ of Eq.~(\ref{eq:SM}) over a sufficiently large number $N_s$ of stripes, we obtain the `statistical-mechanical'' spike-based coherence measure $M_s$:
\begin{equation}
M_s =  {\frac {1} {N_s}} \sum_{i=1}^{N_s} M_i.
\end{equation}

We follow $3 \times 10^3$ stripes and get $O_i$, $P_i$, and $M_i$ in each $i$th stripe for the partially-occupied inhibitory case of Fig.~\ref{fig:PO}(b).  The results are shown in Fig.~\ref{fig:SM1}(c). For comparison, we also measure the degree of stochastic spiking coherence seen in the raster plot of Fig.~\ref{fig:PO}(d) for the fully-occupied excitatory case and give the results in Fig.~\ref{fig:SM1}(d). We note a distinct difference in the
average occupation degree $\langle O_i \rangle$, where $\langle \cdots \rangle$ denotes the average over stripes. For the inhibitory case, partial occupation with very small $\langle O_i \rangle$ $(=0.106)$ occurs due to stochastic spike skipping, while for the excitatory case full occupation with $\langle O_i \rangle =1$ occurs as a result of 1:1 phase locking. For the partially-occupied case of inhibitory coupling, the average pacing degree $\langle P_i \rangle$ $(=0.766)$ is large in contrast to $\langle O_i \rangle$, although it is smaller than $\langle P_i \rangle$ $(=0.911)$ for the fully-occupied case
of excitatory coupling. As a result, the ``statistical-mechanical'' coherence measure $M_s$ (which represents the collective coherence seen in the whole raster plot) is 0.081 for the partially-occupied inhibitory case which is very low when compared with $M_s$ $(= 0.911)$ for the fully-occupied excitatory case. The main reason for the low degree of stochastic spiking coherence is mainly due to partial occupation.
\begin{figure}[t]
\centerline{\includegraphics[width=0.9\columnwidth]{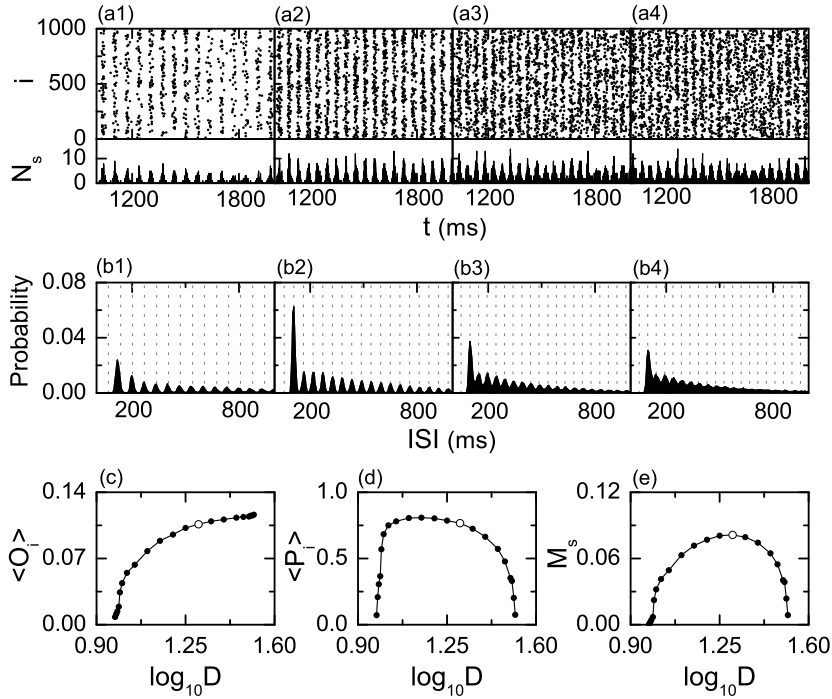}}
\caption{Stochastic spiking coherence for various values of $D$ in $N(=10^3)$ globally coupled inhibitory subthreshold ML neurons when $I_{DC}=87$ $\mu {\rm A /cm^2}$ and $J=3$ ${\rm mS  /cm^2}$. (a) Raster plots of spikings ($i$: neuron index) and plots of number of spikes $(N_s)$ versus time $(t)$ and (b) interspike interval (ISI) histograms for (a1) and (b1) $D=10$ ${\rm \mu A \cdot {ms}^{1/2}/cm^2}$, (a2) and (b2) $D=20$ ${\rm \mu A \cdot {ms}^{1/2}/cm^2}$, (a3) and (b3) $D=30$ ${\rm \mu A \cdot {ms}^{1/2}/cm^2}$, and (a4)
and (b4) $D=32$ ${\rm \mu A \cdot {ms}^{1/2}/cm^2}$. The bin size for $N_s$ in (a) is 1 ms. Each ISI histogram in (b) is made of $5 \times 10^4$ ISIs and the bin size for the histogram is 5 ms. Vertical dotted lines in (b) represent the integer multiples of the global period $T_G$ of $V_G$; $T_G$ = (b1) 67.4 ms, (b2) 54.2 ms, (b3) 48.6 ms, and (b4) 47.7 ms. (c) Plot of $\langle O_i \rangle$ (average occupation degree of spikes) versus $\log_{10} D$. (d) Plot of $\langle P_i \rangle$ (average pacing degree of spikes) versus $\log_{10} D$.
(e) Plot of $M_s$ (``statistical-mechanical'' spike-based coherence measure) versus $\log_{10} D$. To obtain $\langle O_i \rangle$, $\langle P_i \rangle$, and $M_s$ in (c)-(e), we follow the $3 \times 10^3$ stripes for each $D$. Open circles in (c)-(e) denote the data for $D=20$ ${\rm \mu A \cdot {ms}^{1/2}/cm^2}$.}
\label{fig:SM2}
\end{figure}

Finally, by varying $D$ we characterize the stochastic spiking coherence in terms of the ``statistical-mechanical'' measure $M_s$ in a population of inhibitory subthreshold ML neurons when $I_{DC}=87$ and $J=3$. Such stochastic spiking coherence may be well visualized in the raster plots of neural spikes. Figures \ref{fig:SM2}(a1)-\ref{fig:SM2}(a4) show the raster plots and the temporal plots of the number of spikes $(N_s)$ for $N=10^3$ in the coherent region for $D=10$, 20, 30, and 32, respectively. The corresponding ISI
histograms are also given in Figs.~\ref{fig:SM2}(b1)-\ref{fig:SM2}(b4). We measure the degree of stochastic spiking coherence in terms of $\langle O_i \rangle$ (average occupation degree), $\langle P_i \rangle$ (average pacing degree), and $M_s$ for 20 values of D in the coherent regime, and the results are shown in Figs.~\ref{fig:SM2}(c)-\ref{fig:SM2}(e). (Here, for comparison with other values of $D$  we include the case  of $D=20$ which is studied in
details above; open circles in Figs.~\ref{fig:SM2}(c)-\ref{fig:SM2}(e) represent the data for $D=20$.) For the coherent case of D=20, the raster plot in Fig.~\ref{fig:SM2}(a2) consists of relatively clear partially-occupied stripes with $\langle O_i \rangle =0.106$ and $\langle P_i \rangle = 0.766$. Such partial occupation results from stochastic spike skipping of individual
neurons seen well in the ISI histogram of Fig.~\ref{fig:SM2}(b2) with clear (well-separated) multiple peaks appearing at $n\,T_G$ ($T_G$: global period of $V_G$ and $n$=2,3,...). As the value of $D$ is increased from 20, the average occupation degree $\langle O_i \rangle$ increases only a little, as might be seen from the raster plots and the plots of $N_s$ in Figs.~\ref{fig:SM2}(a3) and \ref{fig:SM2}(a4) ($\langle O_i \rangle$ = 0.114 and 0.115 for $D=$ 30 and 32, respectively). This slow increase in $\langle O_i \rangle$ is well shown in Fig.~\ref{fig:SM2}(c). However,the average pacing degree $\langle P_i \rangle$ for $D>20$ decreases rapidly, as shown in Fig.~\ref{fig:SM2}(d). For example, stripes in the raster plots of Figs.~\ref{fig:SM2}(a3) and \ref{fig:SM2}(a4) become more and more smeared, and hence the average pacing degree is decreased with increase in $D$. This smearing of stripes can be understood from the change in the structure of the ISI histograms. As $D$ is increased, the heights of peaks are decreased, but their widths are widened [see Figs.~\ref{fig:SM2}(b3) and \ref{fig:SM2}(b4)]. Thus, peaks begin to merge. This merging of peaks results in the smearing of stripes. We note that for large $D$ merged multiple peaks in the ISI histogram are directly associated with smeared partially-occupied stripes in the raster plot. Thus, for $D>20$ the degree of stochastic spiking coherence is rapidly decreased as shown in Fig.~\ref{fig:SM2}(e), mainly due to the rapid decrease in $\langle P_i \rangle$ (a little increase in $\langle O_i \rangle$ has negligible effect).
Eventually, when passing the higher threshold $D^*_h$ $(\simeq 33.4)$ stripes no longer exist due to complete smearing, and then incoherent states appear.

By decreasing the value of $D$ from 20, we also characterize the stochastic spiking coherence. As an example see the raster plot of spikes in Fig.~\ref{fig:SM2}(a1) for $D=10$. The average occupation degree is much decreased to $\langle O_i \rangle=0.044$, as can also be expected from the plot of $N_s$. This rapid decrease in $\langle O_i \rangle$ for $D<20$ can be seen in Fig.~\ref{fig:SM2}(c). Contrary to $\langle O_i \rangle$ the average pacing
degree $\langle P_i \rangle$ $(=0.684)$ for $D=10$ is decreased a little when compared to the value of $\langle P_i \rangle$ $(=0.766)$ for $D=20$; only a little more smearing occurs. For this case, the ISI histogram has multiple peaks without merging like the case of D=20, as shown in  Fig.~\ref{fig:SM2}(b1). However, when compared to the case of $D=20$ the heights of peaks decrease, and the average ISI increases via appearance of long ISIs. Thus, the degree of stochastic spiking coherence for $D=10$ $(M_s=0.032)$ is much decreased mainly due to rapid decrease in the average occupation degree $\langle O_i \rangle$ (small decrease in $\langle P_i \rangle$ has only a little effect). In fact, as can be seen in Fig.~\ref{fig:SM2}(d) there is no noticeable change in $\langle P_i \rangle$ for $10 < D < 20$; as $D$ is decreased from 20 to a value of $D$ $(\simeq 14)$, $\langle P_i \rangle$ increases slowly, and then it begins to decrease. Thus, for $D<10$ both $\langle O_i \rangle$ and $\langle P_i \rangle$ decreases so rapidly, as shown in Figs.~\ref{fig:SM2}(c) and \ref{fig:SM2}(d). Hence, as $D$ is decreased from 10 the degree of stochastic spiking coherence decreases rapidly due to both effects of $\langle O_i \rangle$ and $\langle P_i \rangle$. Eventually, when $D$ is decreased through the lower threshold $D^*_l$ $(\simeq 9.4)$,
completely scattered sparse spikes appear without forming any stripes in the raster plot, and thus incoherent states exist for $D<D^*_l$. In this way, we characterize the stochastic spiking coherence in terms of the ``statistical-mechanical'' measure $M_s$ in the whole coherent region, and find that $M_s$ reflects the degree of collective coherence seen in the raster plot very well. When taking into consideration both the occupation degree and
the pacing degree of spikes in the raster plot, a maximal spiking coherence occurs near $D \simeq 20$ [see Fig.~\ref{fig:SM2}(e)]. As discussed in details in Section \ref{sec:INT}, $M_s$ is in contrast to the conventional measures where any quantitative relation between (microscopic) individual spikes and the (macroscopic) global potential $V_G$ is not considered [e.g., the ``thermodynamic'' order parameter \citep{GR,HM}, the ``microscopic'' correlation-based measure \citep{WB,White}, and the PSTH-based measures of precise spike timing \citep{PSTH1,PSTH2}].

\section{Summary}
\label{sec:SUM}

By changing the noise intensity, we have characterized stochastic spiking coherence in an inhibitory population of subthreshold biological ML neurons. In a range of intermediate noise intensity, a regularly oscillating global potential $V_G$ (with reduced amplitude and the increased frequency) emerges via cooperation of irregular individual firing activities. We note that this stochastic spiking coherence may be well visualized in the raster plot of spikes. For the case of a coherent state, partially-occupied stripes appear in the raster plot. This partial occupation occurs due to stochastic spike skipping, which is shown clearly in the multi-peaked ISI histogram.
Recently, Brunel and Hakim (1999) have obtained similar results on spike skipping in a simplified network of integrate-and-fire (IF) neurons which are randomly connected via instantaneous inhibitory synapses (modeled by the delta function with a transmission delay). [Such skipping was also reported in the network of inhibitory and excitatory IF neurons \citep{Brunel2,Brunel3}.]
For the case of the IF neuron a spike is triggered instantaneously when the
membrane potential reaches a threshold, in contrast to the case of the biological ML neuron where dynamics of voltage-gated ion channels leads to spike generation. As the stimulus exceeds a threshold, the firing frequency of the IF neuron begins to increase from zero (i.e., the IF neuron exhibits the type-I excitability), unlike the case of the type-II ML neuron (used in our computational study). These type-I IF neurons are driven by random excitatory inputs activated by independent Poisson processes. In this simplified network, it was shown that weakly coherent global activities (roughly corresponding to the global potential $V_G$) emerge from irregular firing patterns of neurons [refer to Figure 3 in the paper of Brunel and Hakim (1999)]. The collective coherence was well shown in the temporal auto-correlation function of the global activity. Although the basic results in the simplified network are similar to ours in a realistic network, there are some differences in the spiking pattern of individual neurons. The stochastic spike emission of individual IF neurons was well shown in the Poissonian ISI histogram showing little indication of the collective behavior. Since no clear multiple peaks appear in the Poisson-like ISI histogram, the individual IF neurons exhibit more stochastic spike emission than the ML neurons, although weak collective coherence occurs in both cases. There are some additional differences in the effect of the external noise on the collective coherence. As the magnitude of the external noise is increased from zero and passes a threshold, the global activity was found to exhibit a transition from an oscillatory to a stationary state \citep{Brunel1}. Thus, incoherent states appear for the case of strong noise. On the other hand, through competition between the constructive role (stimulating coherence between noise-induced spikes) and the destructive role (spoiling the collective coherence) of noise, stochastic spiking coherence (in our work) appears in a range of intermediate noise intensities. Hence, incoherent states appear for both cases of sufficiently strong and weak noise. These differences in both the spiking pattern of individual neurons and the noise effect on the collective coherence seem to arise mainly from different types of drivings on individual neurons; IF neurons are driven by random excitatory inputs activated by Poisson processes, while ML neurons are driven by a Gaussian white noise in addition to a subthreshold DC stimulus.
The main purpose of our work is to quantitatively measure the degree of stochastic spiking coherence seen in the raster plot. We have introduced a new type of spike-based coherence measure $M_s$ by taking into consideration the occupation degree and the pacing degree of spikes in the stripes. Particularly, the pacing degree between spikes is determined in a statistical-mechanical way by quantifying the average contribution of (microscopic) individual spikes to the (macroscopic) global potential $V_G$. This ``statistical-mechanical'' measure $M_s$ is in contrast to the conventional measures (e.g., the ``thermodynamic'' order parameter, the ``microscopic'' correlation-based measure, and the PSTH-based measures of precise spike timing) where any quantitative relation between the microscopic individual spikes and the macroscopic global potential $V_G$ is not considered. In terms of $M_s$, we have quantitatively characterized the stochastic spiking coherence, and found that $M_s$ reflects the degree of collective spiking coherence seen in the raster plot very well. Finally, we also expect that $M_s$ may be implemented to characterize the degree of collective spiking coherence in an experimentally-obtained raster plot of neural spikes.

\begin{acknowledgements}
This research was supported by the Basic Science Research Program through the National Research Foundation of Korea funded by the Ministry of Education, Science and Technology (2009-0070865). S.-Y. Kim thanks  Dr. D.-G. Hong for his interest in our work.
\end{acknowledgements}

\end{document}